\documentstyle[12pt]{article}

\advance \textheight by \topskip
\begin{document}
\begin{flushright}
{\large \bf INR--917/96\\
hep-ph/9602425\\}
November 1995\\
\end{flushright}

\begin{center}
{\large \bf Gauge Dependence of Four-Fermion QED
Green Function and  a Breakdown of Gauge Invariance
in Atom-Like Bound State Calculations}
\vskip0.3cm
{\large Grigorii B. Pivovarov\footnote{e-mail address:
gbpivo@ms2.inr.ac.ru}

Institute for Nuclear Research \\
of the Russian Academy of Sciences,
117312 Moscow, Russia}
\vskip0.3cm
{\large \bf Abstract}
\end{center}
We derive a relation between four-fermion QED Green functions of 
different covariant gauges which defines the gauge dependence 
completely.  We use the derived gauge dependence to check the
gauge invariance of atom-like bound state calculations. We find that 
the existing QED procedure does not provide gauge invariant binding 
energies.  A way to a corrected gauge invariant procedure is pointed 
out.  

\section{Introduction}

QED gives a successful description of atom-like bound states.
The recent measurement of the positronium  life-time \cite{jap}
seems to remove the only discrepancy between theory and 
experiment in this field. Still, one can 
scrutinize general basis of the existing theory which
involves  far from trivial assumptions. The main one is
that excited states correspond to simple poles
of four-fermion QED Green function \cite{Lepage78,Remiddi,Steinman}.
In fact, one cannot prove it because of instability of
excited states.
Next, more technical, is that Bethe-Salpeter kernel is regular
in total energy of fermions near the poles. Combination
of the above assumptions leads to the generally accepted
prescriptions (see, for example, \cite{Lepage78}) for 
calculation of bound state  parameters. Needless to say, any numerical
success yielded by these rules supports but cannot prove the
assumptions.

Let us explain why it is doubtful that the above assumptions
hold. To this end, consider propagator of a charged particle.
Naively, one would expect that it has a simple pole at the
particle mass. But it is well
known (see, for example \cite{Lifshits}) that radiation
of massless photons causes branch point singularity instead of
the simple pole. One should expect the similar effect
for atom-like bound states. The only difference is that two-particle 
bound state
is a dipole. Consequently, one expects the radiation to
be less important. This expectation is in accord with
the successes of the standard approach to the atom-like bound
states.

The aim of this paper is to demonstrate that the main assumption---
correspondence of excited states to simple poles of the Green
function---is in contradiction with gauge invariance.
More precisely, we will show that the assumption leads
to a gauge dependence in the pole positions, i.e., in observable
energy level shifts. We will estimate the leading contribution
to the derivative of level shifts over gauge-fixing parameter.
It will turn out that the gauge dependence is too weak to be seen
in calculations performed up to now.

It may seem that there is an opposite statement in the 
literature. Namely, it was found in \cite{new} that level 
shifts of the standard procedure are gauge invariant.
The difference between \cite{new} and the present paper
lies in the assumptions on the Green function properties 
which were used in the study of gauge invariance. 
In fact, derivation of \cite{new} is based on the above assumptions
which we do not use in our analysis. To be
specific, in the first of two papers
\cite{new} it was pointed out that derivatives of
the Green function over gauge parameter contain branch
points in the total energy of the pair. The authors conclude,
seemingly using the  assumption that the only relevant singularities
are simple poles,
that these branch points should be shifted from the poles corresponding
to the bound states. 
In the present work we allow the possibility
that the Green function have branch points 
and simple poles of coinciding positions. 
In the second paper of \cite{new}, an explicit form
of the level shifts was used to prove their gauge invariance.
The derivation is algebraic in nature and employs implicitly the second
assumption---namely, that Bethe-Salpeter kernels and their
energy derivatives are finite at the poles.
(In notations of \cite{new}, that means finiteness of
$k^{(i)},\dot{k}^{(i)},\ddot{k}^{(i)},...$) 
Again,
we do not use any assumption on the Bethe-Salpeter kernels
in our work (in fact, we even don't need these objects) 
and arrive at an opposite result. 
One may conclude that some of the quantities
$k^{(i)},\dot{k}^{(i)},\ddot{k}^{(i)},...$ 
of \cite{new} are ill-defined. Indeed, more close analysis
proves \cite{Pnew}, that, say, $k^{(5)}$ is infra-red divergent.
We should stress that one would run into these singularities
in the Bethe-Salpeter kernels only in a calculation
of level shifts of order $\alpha^{11}$.

The latter may give a wrong impression that one can safely use 
the standard prescriptions for level shift calculations
up to order $\alpha^{11}$. The real range of applicability of
the standard prescriptions can be found only from a comparison
with new, corrected prescriptions. We  have not them in our 
possession. So, the only claim of the present paper is
that the standard prescriptions break down in order $\alpha^{11}$.

We should anticipate a question on the gauge dependence
of the ground level shift which follows from our general
formulas. Indeed, there is no doubt that ground level of bound system,
if it exists, corresponds to a simple pole of the Green
function. But since
perturbations mix it with excited states, the lack of consistent
picture for exited states causes inconsistency in its description
as well.

The last reservation we should make is on the dependence
of the effect under consideration on the masses of bounded particles.
To simplify the interim formulae, we consider only fermion-antifermion
bound states. But all can be generalized for arbitrary
mass ratio. The mass in the final formulae becomes then 
the reduced mass of the pair. Thus, we claim that even
for the case of infinite mass of the heavier particle, i.e.,
when it can be replaced by the external Coulomb field, the 
effect survives. We expect
that this case may be
the most appropriate one to try to develop new, corrected
prescriptions for level shift calculations.

Turning to a description of the present work itself,
its main technical means is an explicit form of gauge dependence
of the four-fermion QED Green function. We found a relation
between the Green functions of different covariant gauges
which defines the gauge dependence completely. The derivation
is nonperturbative and the relation may present some
interest in itself. It turns out that the gauge dependence 
has a simple form in the space-time representation. To use it,
we formulate a procedure of extraction of level shifts form
the Green function in $x$-representation. Comparison of the 
gauge dependence of the Green function with the extraction
procedure allows us to find the gauge dependence of the level
shifts. We conclude pointing out a possible way to a corrected
gauge invariant procedure of level shift calculations.

Next section contains a derivation of the evolution in the gauge-fixing
parameter; section 3 comprises a brief recall of the extraction procedure
and an utilization of the general evolution formula from section 2
for an analysis of gauge-dependence of the extraction;
in the last, fourth, section we point out the reason for the gauge
dependence and a way to the correct procedure.

\section{Evolution in Gauge-Fixing Parameter}

Let us consider the four-fermion QED Green function
\begin{equation}
\label{Gf}
G_{\beta}(x_{f},\overline{x}_{f},x_{i},\overline{x}_{i})\equiv
        i\int D\psi DA\, \exp\left(iS_{QED}(\beta)\right)
        (\overline{\psi}(\overline{x}_{f}) \psi(x_{f}))
        (\overline{\psi}({x}_{i}) \psi(\overline{x}_{i}))\, ,
\end{equation}
where $x_{f}(\overline{x}_{f})$ is a coordinate of outgoing
particle (antiparticle) and $x_{i}(\overline{x}_{i})$ is the
same for ingoing pair. The definition of
gauge-fixing parameter $\beta$ is given by corresponding photon
propagator:
\begin{equation}
\label{gfix}
D_{\mu \nu}(\beta,x) = \int\frac{dk}{(2\pi)^{4}}
    \left(-g_{\mu \nu} + \beta\frac{k_{\mu}k_{\nu}}{k^{2}}\right)
        \frac{i}{k^{2}}e^{ikx}.
\end{equation}

Our aim is to study the dependence of $G_{\beta}$ on $\beta$.
To this end, it is useful to consider a Green function
in external photon field, $G(A)$, which is a
result of integration over the fermion field in the rhs of 
eq. (\ref{Gf}).
From the one hand, it is simply connected to the Green function:
\begin{equation}
\label{connection}
G_{\beta} = (e^{L_{\beta}}G(A))_{A=0}\,,\;
L_{\beta}\equiv\frac{1}{2}\frac{\delta}{\delta A_{\mu}}D_{\mu \nu}(\beta)
                       \frac{\delta}{\delta A_{\nu}}.
\end{equation}
(In this formula each $L_{\beta}$ generates a photon propagator;
the dependence
on the coordinates of ingoing and outgoing particles is suppressed
for brevity.)
From the other hand, $G(A)$ is simply connected to a gauge invariant 
object $G_{inv}(A)$:  \begin{equation}
\label{coninv} G(A) = G_{inv}(A) 
\exp\left(ie\int^{x_{f}}_{\overline{x}_{f}}A_{\mu}dx^{\mu}
                       -ie\int^{x_{i}}_{\overline{x}_{i}}A_{\mu}dx^{\mu}
                       \right).
\end{equation}
The gauge invariance of $G_{inv}$ means that it is independent of
the longitudinal component of $A$:
\begin{equation}
\label{gi}
\partial_{\mu}\frac{\delta}{\delta A_{\mu}}G_{inv}(A) = 0
\end{equation}
and is a consequence of gauge invariance of the combination
\begin{equation}
\overline\psi(x)\exp\left(ie\int^{x}_{y}A_{\mu}dz^{\mu}
                     \right)\psi(y).
\end{equation}

A substitution of eq. (\ref{coninv}) into eq. (\ref{connection}) yields
\begin{equation}
\label{hot}
G_{\beta} = \left (e^{L_{\beta}}G_{inv}(A)
\exp\left(ie\int^{x_{f}}_{\overline{x}_{f}}A_{\mu}dx^{\mu}
     -ie\int^{x_{i}}_{\overline{x}_{i}}A_{\mu}
dx^{\mu}\right)\right)_{A=0}.
\end{equation}
Let us take a $\beta$-derivative of both sides of this equation:
\begin{equation}
\label{almost eq}
\frac{\partial}{\partial\beta}G_{\beta} =
            \left (e^{L_{\beta}}(\partial_{\beta}L_{\beta})G_{inv}(A)
\exp\left(ie\int^{x_{f}}_{\overline{x}_{f}}A_{\mu}dx^{\mu}
     -ie\int^{x_{i}}_{\overline{x}_{i}}A_{\mu}
dx^{\mu}\right)\right )_{A=0} .
\end{equation}
To get an evolution equation, one needs to express the rhs of
this equation in terms of $G_{\beta}$. It is possible because
$(\partial_{\beta}L_{\beta})$ commutes  with $G_{inv}(A)$ and gives a
$c$-factor when acts on the consequent exponential. So, 
eq. (\ref{almost eq})
transforms itself into
\begin{equation}
\label{equation}
\frac{\partial}{\partial\beta}
G_{\beta}(x_{f},\overline{x}_{f},x_{i},\overline{x}_{i}) =
        F(x_{f},\overline{x}_{f},x_{i},\overline{x}_{i})
        G_{\beta}(x_{f},\overline{x}_{f},x_{i},\overline{x}_{i}),
\end{equation}
where we have restored the $x$-dependence
and used $F$ to denote the action of
$(\partial_{\beta}L_{\beta})$ on the exponential:
\begin{eqnarray}
\label{F-def}
\lefteqn{(\partial_{\beta}L_{\beta})
\exp\left(ie\int^{x_{f}}_{\overline{x}_{f}}A_{\mu}dx^{\mu}
     -ie\int^{x_{i}}_{\overline{x}_{i}}A_{\mu}dx^{\mu}
     \right) \equiv}\nonumber \\
& &       F(x_{f},\overline{x}_{f},x_{i},\overline{x}_{i})
\exp\left(ie\int^{x_{f}}_{\overline{x}_{f}}A_{\mu}dx^{\mu}
     -ie\int^{x_{i}}_{\overline{x}_{i}}A_{\mu}dx^{\mu}
     \right).
\end{eqnarray}

An explanation is in order: In deriving eq. (\ref{equation}) we
have used a commutativity of $(\partial_{\beta}L_{\beta})$
and $G_{inv}(A)$; it is a direct consequence of gauge invariance
of $G_{inv}$ (see eq. (\ref{gi})) and the fact that
$(\partial_{\beta}L_{\beta})$ contains only derivatives in
longitudinal components of $A$ (see eq. (\ref{connection}) for a
definition of $L_{\beta}$ and eq. (\ref{gfix}) for $\beta$-dependence
of $D_{\mu \nu}$).

The solution of eq. (\ref{equation}) for $\beta$-evolution is
\begin{equation}
\label{solution}
G_{\beta}(x_{f},\overline{x}_{f},x_{i},\overline{x}_{i}) =
    \exp\left((\beta-\beta_{0})
    F(x_{f},\overline{x}_{f},x_{i},\overline{x}_{i})
        \right)
        G_{\beta_{0}}(x_{f},\overline{x}_{f},x_{i},\overline{x}_{i}).
\end{equation}

To get the final answer one needs an explicit view of $F$
from eq. (\ref{solution}). It is easily deduced from the $F$-definition
(\ref{F-def}) and the following representation for the longitudinal
part of the photon propagator:
\begin{equation}
\label{representation}
\partial_{\beta}D_{\mu \nu}(\beta,x) =
     -\frac{1}{16\pi^{2}}\partial_{\mu}\partial_{\nu}
     \ln((x^{2}-i\varepsilon)m^{2}),
\end{equation}
where $m$ is an arbitrary mass scale which is fixed, for
definiteness, on the fermion mass. Then, up to an additive constant,
\begin{equation} \label{repres}
F = \frac{\alpha}{4\pi}\left(
 \ln\frac{1}{m^{4}(x_{f}-\overline{x}_{f})^{2}
(x_{i}-\overline{x}_{i})^{2}}
+\ln\frac{(x_{f}-x_{i})^{2}(\overline{x}_{f}-\overline{x}_{i})^{2}}
         {(x_{f}-\overline{x}_{i})^{2}(\overline{x}_{f}-x_{i})^{2}}
                      \right).
\end{equation}

Substituting eq. (\ref{repres}) into eq. (\ref{solution}), we get
our final answer for $\beta$-evolution:  \begin{eqnarray}
\label{answer}
G_{\beta}(x_{f},\overline{x}_{f},x_{i},\overline{x}_{i})&=&
 \left[
      \frac{Z(x_{f}-x_{i})^{2}(\overline{x}_{f}-\overline{x}_{i})^{2}}
           {m^{4}(x_{f}-\overline{x}_{f})^{2}(x_{i}-\overline{x}_{i})^{2}
          (x_{f}-\overline{x}_{i})^{2}(\overline{x}_{f}-x_{i})^{2}}
    \right]^{\frac{\alpha}{4\pi}(\beta-\beta_{0})} \times\nonumber\\
&&G_{\beta_{0}}(x_{f},\overline{x}_{f},x_{i},\overline{x}_{i}) .
\end{eqnarray}
The normalization $Z$ is infinite before the ultraviolet
renormalization. After the renormalization it is scheme-dependent
and calculable order by order in perturbation theory. We
will not need its value in what follows.

\section{The Bound State Parameters And The Four-Fermion QED
Green Function}

The four-fermion QED Green function contains too much
information for one who just going to calculate bound-state parameters.
One can throw away unnecessary information by putting center of mass 
space-time coordinate of ingoing pair and relative times of both 
ingoing and outgoing pairs to zero:  
\begin{equation}
\label{eqtimes} 
G_{(et) \beta}(t,{\bf x},{\bf r'},{\bf r})\equiv
G_{\beta}\left(x_{f}(t,{\bf x},{\bf r'}), \overline{x}_{f}(t,{\bf 
               x},{\bf r'}), x_{i}({\bf r}),
               \overline{x}_{i}({\bf r}) \right),
\end{equation} 
where the space-time coordinates depend on a space-time
coordinate of the center of mass of the outgoing pair $(t,{\bf x})$ 
and a relative space coordinate of outgoing $(\bf r')$ and ingoing 
$(\bf r)$ pair. In the case of equal masses 
\begin{eqnarray}
\label{def r} x_{f}=(t,{\bf x}+\frac{{\bf r'}}{2}),&\;&
\overline{x}_{f}=(t,{\bf x}-\frac{{\bf r'}}{2}),\nonumber \\
x_{i}=(0,\frac{{\bf r}}{2}),&\;&
\overline{x}_{i}=(0,-\frac{{\bf r}}{2}).
\end{eqnarray}

$G_{(et)\beta}$ still contains an unnecessary piece of
information --- the dependence on the center of mass
space coordinate. The natural way to remove it is
to go over to momentum representation and put the
center of mass momentum to zero. In coordinate representation,
which is more convenient for gauge invariance check, we define
the propagator $D_{\beta}$ of the fermion pair:
\begin{equation}
\label{propDef}
G_{(et)\beta}(t,{\bf x},{\bf r'},{\bf r}) \equiv
D_{\beta}(t,{\bf r'},{\bf r})\delta({\bf x}) + \ldots,
\end{equation}
where dots denote terms with derivatives of $\delta({\bf x})$.
It is natural to consider $D_{\beta}$ as a time dependent kernel
of an operator acting on wave-functions of relative coordinate.
In what follows we will not make difference between a kernel
and the corresponding operator. The naturalness of the
above definition of the propagator is apparent in the
nonrelativistic approximation:
\begin{equation}
\label{NR}
{e^{i2mt}}D_{\beta}(t) \approx
                \sum_{E_{0}} \theta(t)e^{-iE_{0}t} P(E_{0}),
\end{equation}
where the summation runs over the spectrum of nonrelativistic
Coulomb problem and $P(E_{0})$ are the projectors onto corresponding
subspaces of the nonrelativistic state space. One can obtain
eq. (\ref{NR}) keeping leading term in $\alpha$-expansion of
the lhs if one will keep $t\propto 1/\alpha^{2}$ and
${\bf r'},{\bf r}\propto 1/\alpha$ (see \cite{Steinman,Pivovarov}).
The subscript on $E_{0}$ is to denote that it will get
radiative corrections (see below). The exponential in the lhs
is to make a natural shift in energy zero.
In what follows we will
include the energy shift in the definition of $
D_{\beta}(t)$.

The next step in calculation of radiative corrections to the energy
levels is a crucial one: one should make an assumption about
the general form of a deformation of the $t$-dependence of
the rhs of eq. (\ref{NR}) caused by relativistic corrections.
A natural guess and the one which leads to the generally accepted
rules of calculation of the relativistic corrections to the energy
eigenvalues (see, for example \cite{Lepage78}) is to suppose that one can
contrive oscillating part of the exact propagator $D_{\beta}$
from the rhs of eq. (\ref{NR}) just shifting energy levels and
modifying the operator coefficients $P(E_{0})$:  
\begin{equation}
\label{guess} D_{\beta}(t) = \sum_{E_{0}+\Delta_{E_{0}}} \theta(t)
                        e^{-i\left(E_{0}+\Delta_{E_{0}}\right)t}
                        P_{\beta}(E_{0}+\Delta_{E_{0}}) + \ldots,
\end{equation}
where dots denote terms which are slowly-varying in time
(the natural time-scale here is $1/E_{0}$). The additional
subscript $\beta$ on $P_{\beta}$ is to denote that
oscillating part of $D_{\beta}(t)$ can acquire a gauge parameter
dependence from relativistic corrections.

The conjecture (\ref{guess}) could be proven if the bound states
were the eigenstates of the Hamiltonian. But being unstable they are
not. We will see that the conjecture (\ref{guess}) contradicts
gauge invariance. Still it turns out extremely
useful---the relativistic corrections calculated with it
are in agreement with the experiment.
Is it possible that another
ansatz may be used instead of eq. (\ref{guess})  preserving its
advantage of success is an open question.

Let us see how one can use eq. (\ref{guess}) in energy level
calculations. It is quite sufficient to consider $D_{\beta}(t)$
on relatively short times when $\Delta_{E_{0}} t\ll 1,\, E_{0}t\sim 1$.
For such times one can approximate $D_{\beta}$
expanding the rhs of eq. (\ref{guess}) over $\Delta_{E_{0}}t$:
\begin{equation}
\label{simple}
D_{\beta}(t) \approx \sum_{E_{0}} \theta(t)e^{-iE_{0}t}
                  \sum_{k}t^{k}A^{(k)}_{\beta}(E_{0}),
\end{equation}
where
\begin{equation}
\label{AE}
A^{(k)}_{\beta}(E_{0}) = \sum_{\Delta_{E_{0}}}
         \frac{(-i\Delta_{E_{0}})^{k}}{k!}P_{\beta}
(E_{0}+\Delta_{E_{0}}).
\end{equation}
An extraction of these objects from the perturbation
theory is an interim step in the level shift calculations.
(Here we should mention that in calculation practice
$A^{(k)}_{\beta}(E_{0})$ are
extracted in momentum representation --- i.e.  not as coefficients
near the powers of time but as the ones near the propagator-like
singularities $(E-E_{0}+i\varepsilon)^{-(k+1)}$.)
To come nearer to the level shift values, useful objects are
\begin{equation}
\label{A}
A^{(k)}_{\beta} \equiv \sum_{E_{0}}A^{(k)}_{\beta}(E_{0})i^{k}k!.
\end{equation}
Namely, as notations of eq. (\ref{AE}) suggest, eigenvalues
of $A^{(0)}_{\beta}$ should be equal to normalizations of bound state
wave functions which are driven from unit by relativistic
corrections while the eigenvalues of $A^{(k)}_{\beta}$ should
be energy shifts to the $k$-th power times corresponding normalizations.
Thus, the eigenvalues of
\begin{equation}
\label{Skdef}
S^{(k)}_{\beta} \equiv \frac{\left[A^{(0)}_{\beta}\right]^{-1}
A^{(k)}_{\beta}
                + A^{(k)}_{\beta}\left[A^{(0)}_{\beta}\right]^{-1}}{2}
\end{equation}
should be just energy shifts to the $k$-th power. Thus, we define
\begin{equation}
\label{Sdef}
S_{\beta} \equiv S_{\beta}^{(1)}
\end{equation}
to be the energy shift operator: its eigenvalues are the energy
level shifts caused by relativistic corrections.
Our aim is now to check $\beta$-independence of $S_{\beta}$
eigenvalues.

Some notes are in order: If the conjecture (\ref{guess})
is true $A^{(0)}_{\beta}$ should commute with $S^{(k)}_{\beta}$
and the following relation should hold:
\begin{equation}
\label{powerrel}
S^{(k)}_{\beta} = \left[S_{\beta}\right]^{k}.
\end{equation}
We will use it in what follows. Another thing to note is
that relativistic corrections affect the form
of the scalar product of wave functions and, thus, one
should add a definition of operator products to the formal expressions 
(\ref{Skdef}),(\ref{powerrel}).  But the level of accuracy to which we 
will operate permits us not to go into this complication and use
the operator products as they are in the nonrelativistic approximation 
--- i.e. as the convolution of the corresponding kernels.

The way to the gauge invariance check of the energy shift calculations
is clear now: Using the gauge evolution relation (\ref{answer})
one should find the $\beta$-dependence of $S_{\beta}$ and then
of its eigenvalues. As $S_{\beta}$ is defined
in eqs. (\ref{Sdef}),(\ref{Skdef}) through $A^{(k)}_{\beta}$'s which are,
in turn, defined in eq. (\ref{simple})
through the propagator $D_{\beta}$, the first step is to simplify
eq. (\ref{answer}) to the reduced case of zero relative time and
total momentum of the fermion pair:
\begin{eqnarray}
\label{reduced}
D_{\beta}(t,{\bf r'},{\bf r})&=&\left[
                        \frac{\left(1-({\bf r'}-{\bf r})^{2}/(4t^{2})
                             \right)}
                            {\left(1-(({\bf r'}+{\bf r})^{2}/(4t^{2})
                             \right)}
                              \right]^
               {\frac{\alpha}{2\pi}(\beta-\beta_{0})}\times\nonumber \\
                             & &\left[
                        \frac{Z}
                            {m^{2}{\bf r'}^{2}m^{2}{\bf r}^{2}}
                              \right]^{\frac{\alpha}{4\pi}
(\beta-\beta_{0})}
                             D_{\beta_{0}}(t,{\bf r'},{\bf r}).
\end{eqnarray}
The factor in the square brackets of the second line is time-independent
and further factorizable on factors depending on either ingoing or 
outgoing pair parameters. This reduce the influence of this factor to 
a change in the normalization of states. Being interested in gauge 
invariance of energy shifts, we omit this factor in what follows.  Let 
us turn to the analysis of the influence of the factor in the first
line of eq. (\ref{reduced}).

This factor is close to unit in the atomic scale
${\bf r'},{\bf r}\sim 1/\alpha,\,t\sim1/\alpha^{2}$.
We will use its approximate form:
\begin{equation}
\label{approx}
Factor \approx 1 + \frac{\alpha}{2\pi}(\beta-\beta_{0})
                \frac{{\bf r'}{\bf r}}{t^2} + O(\alpha^{5}).
\end{equation}

One can read the dependence of $A^{(k)}_{\beta}$
on $\beta$ from eqs. (\ref{simple}),(\ref{reduced}),(\ref{approx})
as
\begin{equation}
\label{betadep}
A^{(k)}_{\beta} \approx A^{(k)}_{\beta_{0}} -
               \frac{\alpha}{2\pi}\frac{(\beta-\beta_{0})}{(k+1)(k+2)}
               {\bf r}A^{(k+2)}_{\beta_{0}}{\bf r},
\end{equation}
where $\bf r$ is the vector operator of relative position
of interacting particles.
The mixing of different $A^{(k)}_{\beta}$'s with a change in
the gauge parameter is due to the presence of $1/t^{2}$ in
the rhs of eq. (\ref{approx}).
Finally, using the definition
(\ref{Sdef}), relations (\ref{powerrel}) and the fact that
\begin{equation}
\label{unit}
A^{(0)} \approx 1
\end{equation}
in the nonrelativistic approximation one can derive the
following $\beta$-dependence of $S_{\beta}$:
\begin{eqnarray}
\label{Sanswer}
S_{\beta}&\approx&S_{\beta_{0}} -\nonumber \\
         &       &\frac{\alpha}{2\pi}(\beta-\beta_{0})
                  \left(\frac{1}{6}{\bf r}S_{\beta_{0}}^{3}{\bf r} -
                \frac{1}{4}S_{\beta_{0}}{\bf r}S_{\beta_{0}}^{2}{\bf r} -
                 \frac{1}{4}{\bf r}S_{\beta_{0}}^{2}{\bf r}S_{\beta_{0}}
                   \right).
\end{eqnarray}
Treating the term in the last line of the rhs of the above relation
as a perturbation, one can get an approximate
value of the $\beta$-dependent
piece of the energy shift just averaging the perturbation
with respect to the corresponding eigenstate of $S_{\beta_{0}}$.

Thus, we get for the leading order of $\beta $-derivative of
an energy shift the following representation:
\begin{equation}
\label{leading}
\left(\frac{\partial}{\partial\beta}\Delta_{\beta}\right)_{L}=
               -\frac{\alpha}{2\pi}
                  \left(\frac{1}{6}\left\langle
                  {\bf r}S_{L}^{3}{\bf r}\right\rangle -
  \frac{1}{4}\left\langle S_{L}{\bf r}S_{L}^{2}{\bf r}\right\rangle -
  \frac{1}{4}\left\langle{\bf r}S_{L}^{2}{\bf r}S_{L}\right\rangle
                   \right),
\end{equation}
where $\langle\ldots\rangle$ means averaging with respect to
the corresponding nonrelativistic eigenstate and the subscript
$L$ means the leading  order in $\alpha$-expansion.

Eq. (\ref{leading}) is sufficient to define an order in
$\alpha$ in which the energy shifts become gauge dependent:
\begin{equation}
\label{order}
\left(\frac{\partial}{\partial\beta}\Delta_{\beta}\right)_{L}
                        \sim \alpha^{11}.
\end{equation}
Here we have taken into account that ${\bf r}\sim1/\alpha$
and $S_{L}\sim\alpha^{4}$.

To have a gauge dependence in any observable is clearly
unacceptable. In the next section we will see how one
should correct the above procedure of energy shift
extraction from the QED Green function to get rid of
the gauge dependence of energy shifts.

\section{A Way Out}

The procedure recalled in the previous section
is based on the conjecture (\ref{guess}).
A consequence of this conjecture is the
gauge dependence of energy shifts of eq. (\ref{leading}).
One can conclude that the conjecture is wrong.
In particular, as one can infer from eq. (\ref{reduced}),
the operator coefficients near the oscillating exponentials
in eq. (\ref{guess}) should get a time dependence from relativistic
corrections. Even if in some gauge they are time independent,
the gauge parameter evolution should generate a dependence
which in the leading order in $\alpha$ reduce itself to
the following replacement in eq. (\ref{guess}):
\begin{equation}
\label{replacement}
P_{\beta}(E_{0}+\Delta_{E_{0}})\rightarrow
P_{\beta}(E_{0}+\Delta_{E_{0}}) + \frac{\Sigma_{\beta}(E_{0})}{t^{2}}.
\end{equation}
That $\Sigma_{\beta}(E_{0})$ has nothing to do with
energy shifts but will give contributions
to $A^{(k)}_{\beta}(E_{0})$'s from eq. (\ref{simple}).
Being gauge dependent these contributions lead to the
gauge dependence of energy shifts.

The way to the correct procedure is to throw away
terms like $\Sigma_{\beta}(E_{0})/t^{2}$ prior to the definition of
the energy shift operator. Thus, a necessary step
in the process of extracting energy shifts from
the QED Green function (and the one
which necessity is not recognized in the standard procedure) is to 
calculate and subtract contributions like the last term in the rhs of 
eq. (\ref{replacement}) from the propagator of the fermion pair.

Below we report on a calculation of $\Sigma_{\beta}(E_{0})$ from
eq. (\ref{replacement}). The most economical way to calculate it
is to note that the energy dependence
of the Fourier transform of the corresponding
contribution to the propagator is
\begin{equation}
\label{fourier}
(E-E_{0})\ln(-(E-E_{0}+i\varepsilon))
\end{equation}
and that it comes from diagrams describing radiation
and subsequent absorption of a soft photon with no change
in the level $E_{0}$ of the radiating and absorbing bound state.
Similar contributions (with another power of energy before the $log$)
are well known for the propagator of a charged fermion \cite{Lifshits}.

It may be worth to note here that contribution of eq. (\ref{fourier})
vanishes at $E=E_0$. This explains why such contributions
are insignificant for practical calculations of the present day
accuracy. In particular, one can neglect them, despite the 
log-singularity,
in the resonance scattering calculations and preserve the classic
results of \cite{low}.

The first step in our calculation is to
present the pair propagator in the following form:
\begin{equation}
\label{soft}
D_{\beta}(t)\approx\left(e^{L_{s}}e^{ie{\bf rA}(t)}D_{inv}(t,A)
                       e^{-ie{\bf rA}(0)}
               \right)_{A=0},
\end{equation}
where $L_{s}$ is the same as in eq. (\ref{connection}) except a
restriction on the momentum of photon propagator ---
the range of its variation is restricted to the soft
region which border is of order of atomic
binding energies; the exponentials with
gauge potential are originated from
the ones in eq. (\ref{hot}); $D_{inv}$ is a descendant of $G_{inv}$
from (\ref{hot}): to go over from $G_{inv}$ to $D_{inv}$
one should make all pairing of non-soft photons in $G_{inv}$
and all the reductions of space-time coordinates which was involved in 
going over from the $G_{\beta}$ of eq. (\ref{Gf}) to the $D_{\beta}$
of eq. (\ref{propDef}); at last, all gauge potentials in eq. (\ref{soft})
are taken at zero of space coordinate in accord with the
$\delta({\bf x})$ of eq. (\ref{propDef}). The difference
between the lhs and the rhs  of eq. (\ref{soft}) does not contribute
to the term under the calculation.

The leading in the nonrelativistic approximation contribution to
$D_{inv}$ is the same as for $D_{\beta}$ --- it is just
the propagator of the nonrelativistic Coulomb problem. We explicitly
calculate the leading contribution
to the dependence of $D_{inv}(t,A)$ on the gauge potential
in its expansion over soft momenta of the external photons.
Not surprisingly, the dipole interaction of the pair with
the external photon field arises in this approximation:
\begin{equation}
\label{Adef}
D_{inv}(t,A) \approx \left(i\frac{\partial}{\partial t} - H_{c}
                                        + e{\bf r}{\cal E}(t)
                  \right)^{-1},
\end{equation}
where $H_{c}$ is the Hamiltonian of the nonrelativistic Coulomb 
problem and $\cal E$ is the strength of the electric field:
\begin{equation}
\label{Edef}
{\cal E}(t)\equiv -\dot{{\bf A}}(t) + \nabla A_{0}(t).
\end{equation}

Substituting eq. (\ref{Adef}) in eq. (\ref{soft}) and keeping terms
with only one soft photon propagator we get
expressions which sum contains the term under calculation:
\begin{equation}
\label{r1}
e^{2}\left(L_{s}
                {\bf rA}(t)D_{nr}(t){\bf rA}(0)\right)_{A=0},
\end{equation}
\begin{equation}
\label{r2}
e^{2}\left(L_{s}
                \int d\tau_{1}d\tau_{2}\,
                D_{nr}(t-\tau_{1}){\bf r}{\cal E}(\tau_{1})
                D_{nr}(\tau_{1}-\tau_{2}){\bf r}{\cal E}(\tau_{2})
                D_{nr}(\tau_{2})\right)_{A=0} ,
\end{equation}
\begin{eqnarray}
\label{r3}
ie^{2}\biggl(L_{s}
                \int d\tau\,\bigl(
          D_{nr}(t-\tau){\bf r}{\cal E}(\tau)D_{nr}(\tau){\bf rA}(0)&-&
          \\
&  &      {\bf rA}(t)D_{nr}(t-\tau){\bf r}{\cal E}(\tau)D_{nr}(\tau)
          \bigr)
                       \biggr)_{A=0},\nonumber
\end{eqnarray}
where $D_{nr}(t)$ is the propagator of the nonrelativistic Coulomb
problem from the rhs of eq. (\ref{NR}).

The next step is to pick out a contribution of a level $E_{0}$
in eqs. (\ref{r1}),(\ref{r2}),(\ref{r3}). That is achievable by the
replacement
\begin{equation}
\label{repl}
D_{nr}(t)\rightarrow e^{-iE_{0}t}\theta(t)P(E_{0}).
\end{equation}

The last ingredient that one needs to calculate eqs.
(\ref{r1}),(\ref{r2}),(\ref{r3}) is the time dependence of the soft
photon propagators. It can be deduced from eq. (\ref{gfix})
as
\begin{eqnarray}
\label{time}
\left(L_{s}A_{i}(t_{1})A_{j}(t_{2})\right)&=&
                \theta\left((t_{1}-t_{2})^{2}>t_{c}^{2}\right)
 \frac{\delta_{ij}\left(-1+\frac{\beta}{2}\right)}
{4\pi^{2}(t_{1}-t_{2})^{2}},
 \nonumber \\
\left(L_{s}A_{i}(t_{1}){\cal E}_{j}(t_{2})\right)&=&
                \theta\left((t_{1}-t_{2})^{2}>t_{c}^{2}\right)
 \frac{\delta_{ij}}{2\pi^{2}(t_{1}-t_{2})^{3}},\nonumber \\
\left(L_{s}{\cal E}_{i}(t_{1}){\cal E}_{j}(t_{2})\right)&=&
                \theta\left((t_{1}-t_{2})^{2}>t_{c}^{2}\right)
 \frac{\delta_{ij}}{\pi^{2}(t_{1}-t_{2})^{4}}.
\end{eqnarray}
Here the $\theta$-functions are to account for the softness of
the participating photons ($t_{c}\sim 1/E_{0}$).

Taking eq. (\ref{time}) into account we get the
following contributions from
eqs. (\ref{r1}),(\ref{r2}),(\ref{r3}):
\begin{eqnarray}
\label{contr}
(\ref{r1})&\rightarrow& \frac{1}{t^{2}}\theta(t)e^{-iE_{0}t}
        \frac{\alpha}{\pi}\left(-1+\frac{\beta}{2}\right)
        {\bf r}P(E_{0}){\bf r},\nonumber \\
(\ref{r2})&\rightarrow& \frac{1}{t^{2}}\theta(t)e^{-iE_{0}t}
  \frac{\alpha}{\pi}\frac{2}{3}P(E_{0}){\bf r}P(E_{0}){\bf r}P(E_{0}),
  \nonumber \\
(\ref{r3})&\rightarrow& \frac{1}{t^{2}}\theta(t)e^{-iE_{0}t}
  \frac{\alpha}{\pi}i\left(P(E_{0}){\bf r}P(E_{0}){\bf r} -
  {\bf r}P(E_{0}){\bf r}P(E_{0})\right).
\end{eqnarray}

The sum of the above terms yields the result of our calculation:
\begin{eqnarray}
\label{sigmansw}
\Sigma_{\beta}(E_{0})&=&\frac{\alpha}{\pi}
\biggl( \frac{2}{3}P(E_{0}){\bf r}P(E_{0}){\bf r}P(E_{0}) +
(-1+\frac{\beta}{2}){\bf r}P(E_{0}){\bf r} +\nonumber\\
    & &  i(P(E_{0}){\bf r}P(E_{0}){\bf r} - {\bf r}P(E_{0}){\bf r}
P(E_{0}))
\biggr) .
\end{eqnarray}

One can explicitly check that $\beta$-dependence of 
$\Sigma_{\beta}(E_{0})$
is the right one --- i.e. if one subtracts the $\Sigma$-term
from the propagator before the definition of the energy shift
operator, the latter becomes gauge independent. Another observation
is that the $\Sigma$-term cannot be killed by any choice
of the gauge (in contrast to the case of charged fermion propagator
where an analogous term is equal to zero in the Yennie gauge).

Summing up, in this paper we derived a relation between QED
Green functions of different gauges. We used it to check
the gauge invariance of the energy shift operator. It turns out
to be gauge dependent. This fact forced us to recognize
that energy shifts are not one, and the only one, source for
the positive powers of time near the oscillating exponentials
in the propagator of the pair. We found a particular
additional source of the positive powers of time which is
responsible for the gauge dependence of the naive
energy shift operator. We conclude with an observation
that at the moment we have not a  clear
definition of the energy shift operator --- to get
it one needs a criterion for picking out contributions
to the positive powers of time originating from the energy
shifts.

The author is grateful to A.~Kataev, E.~Kuraev, V.~Kuzmin, A.~Kuznetsov,
S.~Larin, Kh.~Nirov, E.~Remiddi,
V.~Rubakov, D.~Son, P.~Tinyakov
for helpful discussions. This work was supported in
part by Russian Foundation for Basic Research,
project no. 94-02-14428.

\end{document}